\documentclass[graybox]{svmult}
\usepackage{tabu} 

\usepackage{type1cm}        
\usepackage{makeidx}         
\usepackage{graphicx}        
\usepackage{multicol}        
\usepackage[bottom]{footmisc}

\usepackage{newtxtext}       %
\usepackage{newtxmath}       

\makeindex             

\begin{document}

\title*{Mixing dynamics of dimension-five interactions (scalar/pseudoscalar-photon) in magnetized medium}
\titlerunning{Mixing dynamics of Dim-5 interactions } 
\author{Ankur Chaubey, Manoj K. Jaiswal, Avijit K. Ganguly}
\institute{Ankur Chaubey \at Institute of Science, Department of Physics, Banaras Hindu University, Varanasi-India.\\ \email{ankur.chaubey@bhu.ac.in}
\and Manoj K. Jaiswal \at  Department of Physics, University of Allahabad, Prayagraj 211002-India.\\ \email{mkjaiswalbhu@gmail.com }
\and Corresponding author: Avijit K. Ganguly\at Department of Physics(MMV), Banaras Hindu University, Varanasi-India.\\ \email{Avijitkganguly@gmail.com}}
%
\maketitle

\abstract {In many extentions of standard model, dimension-5 scalar di-photon 
($g_{\gamma\gamma\phi}\phi$ $F^{\mu\nu}F_{\mu\nu}$ ) or pseudoscalar di-photon
($g_{\gamma\gamma a}a\tilde{F}^{\mu\nu} F_{\mu\nu} $, ) interaction  
materializes due to scale symmetry breaking or $U_A(1)$ symmetry breaking.
In a magnetized vacuum (i.e., in an external background field $\bar{F}_{\mu\nu} $)
the transverse degrees of freedom of the photons-- for such systems-- can be described
in terms of the form factors constructed out of the   background field strength tensor ($\bar{F}_{\mu\nu} $) and the same for 
dynamical photon ($f^{\mu\nu}$); they happen to be $\bar{F}_{\mu\nu}f^{\mu\nu}$ and 
${\tilde{\bar{F}}_{\mu\nu}f^{\mu\nu}}$. 
These form factors transform differently under CP transformation.  While $\bar{F}_{\mu\nu}f^{\mu\nu}$ 
(describing polarization orthogonal to B ($|\gamma_{||} >$)) is CP even,
 the other one, ${\tilde{\bar{F}}_{\mu\nu}f^{\mu\nu}}$( describing polarization along B 
($|\gamma_\bot >$), is CP odd.  In the interaction Lagrangian, if the scalar is interchanged with 
the pseudoscalar, the role of the two form factors just gets interchanged. Thus  
for nearly degenerate strengths of the 
coupling constants ( $g_{\gamma\gamma\phi}$ and $g_{\gamma\gamma _a}$ ) and masses ($m_\phi$ 
and $m_a$ ) of the respective candidates, proper identification of one from the other may 
become very difficult in laboratory or  astrophysics based experiments.
The basic motivation of this investigation is to reduce this uncertainty 
through incorporation of parity violating  ({\it originating through magnetized medium effects }) 
part of the photon self-energy  in the effective Lagrangian. 
This step, in turn affects the (Pseudo) Scalar Photon  mixing dynamics drastically and
brings out a significant change in the spectrum of the electromagnetic beam undergoing
such interaction.
}

\section{Introduction}
\label{sec:1}
\label{sec:1}
Pseudoscalar particles like axions a(x) are common to be 
associated with the breaking of chiral symmetries
in many theories of unification (in physics beyond the standard model) 
\cite{Peccei,Weinberg,Wilczek,Mohapatra,Kim,donoghue,Witten,
Ashok, Marsh} through quantum effects; 
and so is also the case with the Goldstone bosons of a spontaneously broken scale symmetry (dilaton) $\phi(x)$ \cite{Kim, donoghue}. They both have remained possible candidates of Dark matter for some times now. The interaction dynamics of these exotic particles i.e. scalars$(\phi(x))$ or pseudoscalars$(a(x))$ with photon $(\gamma)$ is governed by  Dim-5 operators $g_{\gamma\gamma\phi}\phi F ^{\mu\nu} F_{\mu\nu}$ or $g_{\gamma\gamma a}a \tilde{F}^{\mu\nu} F_{\mu\nu} $. \\

\indent
The associated form factors $\bar{F}_{\mu\nu}f^{\mu\nu}$ and 
${\tilde{\bar{F}}_{\mu\nu}f^{\mu\nu}}$,
for the transverse degrees of freedom of the photons
-- in external background field $\bar{F}_{\mu\nu} $ --for such systems 
have different CP transformation properties.
As a result, in the equation of motion,  
the CP even form factor $\bar{F}_{\mu\nu}f^{\mu\nu}$ couples only to the 
CP even scalar field $\phi(x)$ while the other CP odd  one, ${\tilde{\bar{F}}_{\mu\nu}f^{\mu\nu}}$, propagates freely.  In other words, out 
of the three available degrees of freedom,
the mixing is between only two degrees of freedom-- having identical CP 
properties \cite{Raffelt}. And most importantly: the mixing matrix is 
$2 \times 2$.
On the other hand for magnetized 
pseudo-scalar photon system reverse happens, i.e. the roles of the 
form factors $\bar{F}_{\mu\nu}f^{\mu\nu}$ and  
${\tilde{\bar{F}}_{\mu\nu}f^{\mu\nu}}$ get interchanged.\\
\indent
Further more, the presence of the external field, compromises the
Lorentz symmetry for both the systems; identically. For an external magnetic
field in the $z$ direction ({\bf $B_z$}), except for 
rotational and boost symmetry around and along  {\bf $B_z$}, all other 
space-time symmetries get compromised. This manifests itself by turning 
the vacuum into an optically active and  dichroic medium, for the 
photons\cite{Raffelt,Miani, Ganguly-jaiswal} passing through such region. 
Utilizing this, standard 
polarimetric observables like polarization or ellipticity  angle
can be measured and used to determine the magnitude of the coupling
constant and mass $g_{\gamma\gamma\phi}$ and $m_{\phi}$ for $\phi-\gamma\gamma$
system or  $g_{\gamma\gamma a}$ and $m_{a}$  for  $a-\gamma\gamma$
system. This process of determination is however subject to cross-correlated 
verification from other experiments  for example  \cite{wilczek}  . \\
\indent
However, the $2\times2$ nature of the  mixing matrix for both $\phi(x)\gamma\gamma$ and $a(x)\gamma\gamma$ \,system poses a problem,  when the magnitude of the  masses $m_{a}$ and $m_{\phi}$ as well as the  coupling constants  ($g_{\gamma\gamma a}$) and 
($g_{\gamma\gamma \phi}$) are close to each other. In such a scenario, the
identification of one from the other is difficult using the polarimetric 
techniques. The reason being, as one moves from  $\phi(x)\gamma\gamma$  
to  $a(x)\gamma\gamma$ system, the role of the two polarization 
form-factors gets interchanged with each other. As a result the absolute 
magnitude of the ellipticity and polarization angle remain same. And the 
degree of polarization also remain insensitive to the underlying theory.\\ 
\indent 
The main motivation of this study is to explore other physical corrections,
such that the incorporation of them  would eventually break the degeneracy 
in the  $2\times 2$  mixing pattern undergone by both $\phi(x)-\gamma\gamma$ 
and  $a(x)-\gamma\gamma$ system in a magnetized vacuum.\\

\indent
It so happens that, as one incorporates the parity-violating part of 
photon-self-energy-tensor (PSET) , that appears once the effect of
magnetized medium is incorporated in the evaluation of PSET, in the 
effective Lagrangian of the system, the apparent degeneracy in mixing 
gets lifted. This happens due to the discrete symmetries enjoyed by the
respective form factors of the photon  as well as the scalar or 
pseudoscalar field.     
With the incorporation of such effect, the mixing matrix for 
$\phi F_{\mu\nu}F^{\mu\nu}$, type of interactions, turns out to be $3\times 3$
and for $a F_{\mu\nu}F^{\mu\nu}$  interaction  the mixing matrix is  
$4 \times 4$. That is there is mixing of all four degrees of freedom --three 
degrees of freedom of the in-medium photon and one degree of freedom of 
the pseudoscalar, for $a(x)-\gamma\gamma$ system.
    
\section{Mixing dynamics of Scalars and Pseudosclars in magnetised plasma }
\label{sec:2}

The action for scalar photon system,, as the quantum corrections due to ambient medium and an external magnetic field
${\bf eB}$ are taken into account \cite{ch2-GKP}, turns out to be
\begin{small}
\begin{eqnarray}
S &=& \int d^4k \left[\frac{1}{2} 
A^\nu(-k) 
 \left( -k^2 \tilde{g}_{\mu\nu} + \Pi_{\mu\nu}(k) + \Pi^{p}_{\mu\nu}(k) \right) A^\mu(k)
\nonumber  \right. \\ \left.
\right. && \left. + ig_{\phi\gamma\gamma}{\phi(-k)}\bar{F}_{\mu\nu}k^\mu A^\nu(k)+\frac{1}{2}\phi(-k)[k^2 - m^2]\phi(k) \right].
\label{action}
\end{eqnarray}
\end{small}
Here $\Pi_{\mu\nu}(k)$ is the in medium polarization tensor and 
$\Pi^{p}_{\mu\nu}(k)$ the parity violating part of the same evaluated 
in a magnetized medium.
One can get the same for pseudoscalar/scalar-photon system from eqn. 
(\ref{action}), by replacing $\phi(\pm k)$ by $a(\pm k)$ and $\bar{F}_{\mu\nu} $ 
by $\tilde{\bar{F}}_{\mu\nu}$. Derivation of the equations of motions
follows next.

\subsection{Mixing matrix of Scalar photon interaction }
The equations of motion,  of scalar photon system follow from 
eqn. (\ref{action}). Written in matrix the same is:
\begin{equation}
\left[\begin{array}{c} k^2 {\bf I} - 
               \left( \begin{array}{ccc}              
       \omega^2_p                           &    i\frac{\omega^{2}_{p} eB_{\parallel} }{(\omega m_e)}     & \,\,   -i g_{\phi\gamma\gamma}B_{\perp}\omega   \\
-i\frac{\omega^{2}_{p} eB_{\parallel} }{(\omega m_e)}    &          \omega^{2}_{p}                        &                      0                     \\
ig_{\phi\gamma\gamma}B_{\perp}\omega        &                       0                         &                    m^2_{\phi}  
   \end{array} \right)
 \end{array} \right]
\left[\begin{array}{c} A_{\parallel}(k) \\ A_{\perp}(k) \\\phi(k) \end{array} \right]=0.
\label{photon-scalar-mixing-matixx4}
\end{equation} 

\noindent
The longitudinal degree of freedom, doesn't couple to anything, it  
propagates freely. The same can be explained with the help of
the discrete symmetries enjoyed by the form factor associated with
the longitudinal degree of freedom of the photon. 
Hence the mixing is between  $A_{\parallel}(k),$  $A_{\perp}(k)$
and $\phi$ only.  Where  $A_{\parallel}(k),$  $A_{\perp}(k)$
are the form factors associated with the 
degrees of freedom of photon those are-- parallel and perpendicular to the
 direction of magnetic field. We had obtained the solutions of eqn.(\ref{photon-scalar-mixing-matixx4}) by diagonalizing the mixing matrix.
\subsection{Mixing matrix of axion photon interaction }
As before the equations of motion for axion photon system, can be 
expressed in matrix notation as, 
\begin{equation}
\left[(\omega^2 + \partial_z^2){\bf I} -{\bf M^{\prime} }\right]
\left( \begin{array}{c}
A_{\parallel}(k)   \\
A_{\perp}(k)   \\
A_{L}(k)  \\
a (k)  \\
\end{array} \right)=0,
\label{axion_field}
\end{equation}
where ${\bf I}$ is an identity matrix and   matrix ${\bf M^{\prime}}$ is 
the  $4\times 4$ mixing matrix. The same, in terms of its
elements is given by,
\begin{equation}
{\bf M^{\prime}} = \left( \begin{array}{cccc}
 \Pi_T & - \Pi_p N_1 N_2 P_{\mu\nu}b^{(1)\mu}I^{\nu} & 0 & 0 \\
\Pi_p N_1 N_2 P_{\mu\nu} b^{(1)\mu}I^{\nu}  &  \Pi_T
& 0 & -ig_{a\gamma\gamma}{N_2b^{(2)}_\mu I^{\mu}} \\
0 & 0 &  {\Pi_L}     &   -ig_{a\gamma\gamma}{N_Lb^{(2)}_\mu \tilde{u}^{\mu}}\\
0 & ig_{a\gamma\gamma}{N_2b^{(2)}_\mu I^{\mu}} &  i
g_{a\gamma\gamma}{N_L b^{(2)}_\mu \tilde{u}^{\mu}}  & m^2_{a}
\end{array} \right).
\label{mat-axion-photon}
\end{equation}
\noindent
We note that, projection operator $P_{\mu\nu}$, appearing in the $M^{\prime}_{12}$ and   $M^{\prime}_{21}$ 
elements of the mixing matrix $M^{\prime}$ is a complex one, that makes the matrix, $M^{\prime}$, a hermitian 
matrix, that is expected even otherwise on general grounds.\\
\indent
It is also important to note that for pseudoscalar-photon interaction, because of discrete symmetry considerations (PT symmetry to be specific )the 
form factor associated with longitudinal degree of freedom remains coupled with pseudoscalar field. Hence the mixing matrix becomes $ 4 \times 4$. 
Therefore the mixing dynamics for these two systems with incorporation of 
parity violating medium effect, turns out to be completely different. 
Due to this, the identification of one from the other using 
polarimetric observables may become lot easier.

\section{Optical observables  }
\label{sec:2}
\noindent
Properties of polarized light waves can be described in terms of the Stokes parameters
evaluated from the coherency matrix. The same is constructed from the solutions of the field equations;
and is given by: \\
\begin{equation}
{\bf D'}(z) =
    \left( \begin{array}{cc}              
       < {\bf A_{\parallel}(\omega,z) A^{*}_{\parallel}(\omega,z)} >    &   < {\bf A_{\parallel}(\omega,z) A^{*}_{\perp}(\omega,z)} >     \\
 < {\bf A_{\perp}(\omega,z) A^{*}_{\parallel}(\omega,z)} >    &   <{\bf A_{\perp}(\omega,z) A^{*}_{\perp}(\omega,z) }>             
   \end{array} \right).
\label{density-mat}   
\end{equation}\\

\noindent
In eqn. (\ref{density-mat}) above, $<  >$ represent the ensemble averages.
The Stokes parameters are obtained from the elements of the coherency matrix by the following
identifications; 
${\bf I} = D'_{11}(z) + D'_{22}(z)$,  ${\bf Q} = D'_{11}(z) - D'_{22}(z)$ ,  
${\bf U} = 2 Re\; D'_{12}(z)$, and ${\bf V} = 2 Im\; D'_{12}(z)$.

The estimates of the other optical parameters, i.e.,  ellipticity angle, polarization angle, degree of 
linear polarization, degree of total polarization, follows from  the expressions of  {\bf I}, {\bf U}, 
{\bf Q} and {\bf V}. The expressions for the polarization angle and ellipticity angle, associated  with an electromagnetic wave are provided 
below.
\subsection{Polarization Angle \& Ellipticity Angle}
Polarization angle (represented by $\Psi$) is the angle between major and minor 
axis of ellipse,  defined  in terms of stokes parameters ${\bf U}$ and ${\bf Q}$, 
is given by, 
\begin{eqnarray}
tan(2\Psi) = \frac{{\bf U}(\omega,z)}{{\bf Q}(\omega,z)}.
\end{eqnarray}
The ellipticity angle (denoted by $\chi$) is defined in terms of the same parameters as,
\begin{eqnarray}
tan(2\chi) = \frac{{\bf V}(\omega,z)}{\sqrt{{\bf Q}^2(\omega,z) + {\bf U}^2(\omega,z)}}.
\end{eqnarray}

\section{Results and Conclusions }
\label{sec:3}
\begin{figure}[ht!]
\begin{center}
\includegraphics[scale=.6]{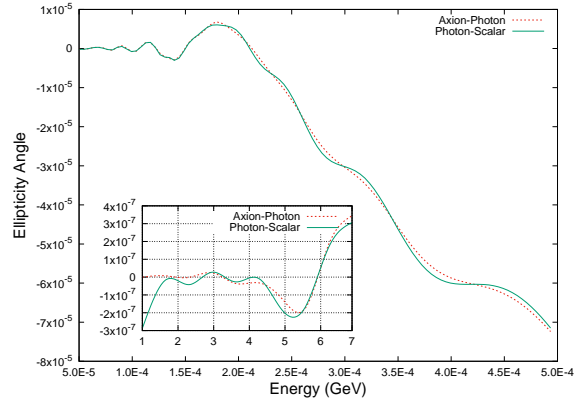}
\caption{\label{ch.52}\small{Plot for ellipcity angle vs energy in case of coupled photon-axion system.
The abscissa of the plot, 
in the inset is, in units of $10^{-5}$ GeV.  }}   
\label{ch.ellipangle5}
\end{center}
\end{figure}

Unlike polarization angle, the ellipticity angle remains invariant under 
rotation of the axes. 
So we have compared the magnitude of the ellipticity angle produced
through  axion-photon as well as scalar-photon interaction,
in the vicinity of a strongly magnetized compact astrophysical source. 
The parameters, that we have considered for the system are as follows:
plasma frequency $\omega_p = 1.6 \times 10^{-10}$ GeV, coupling 
constants $ g_{\gamma\gamma a } = g_{\gamma\gamma\phi}\!=\!10^{-11} GeV^{-1} $
and mass $m_{a}$ and $m_{\phi}$  both close to zero.  
The magnetic field is taken to be $B = 10^{12}$ Gauss and the
 path length considered here is 2.5 Km.. 
The numerical  estimates of the ellipticity angle  for the two systems 
are plotted in, Fig.[\ref{ch.ellipangle5}].\\

\indent
As can be seen in the plot that -- for the values of the  parameters chosen here
--- the numerical magnitudes of the angle
for the $a \gamma$ and  $\phi \gamma$ system, are extremely close to each 
other.  However there is some departure, that can be seen in the inset of 
Fig.[1]. 
In the energy range  of, $1\times 10^{-5}$GeV to $ 1.5\times 10^{-5}$ GeV, 
there is some visible difference in the  ellipticity  angle between 
axion photon and scalar photon systems.
For energies  close $1\times 10^{-5}$ GeV  the difference is around 
$3 \times 10^{-7}$rad.
Though this is little less for current sensitivity available for the 
detectors, however we hope that, future detectors would have similar 
sensitivity to resolve this difference and shed light on the values 
of the parameters like $g_{\gamma\gamma \phi}$ or  $g_{\gamma\gamma a}$ and   
$m_{\phi}$ or $m_{a}$. Studies along this direction are currently
under progress and would be communicated else where shortly.

\end{document}